\documentclass[]{aa} 

\usepackage{txfonts}
\usepackage{graphicx}
\usepackage{natbib}

\newcommand{\CHAX}{CH~A$^{2}\Delta$--X$^{2}\Pi$}
\newcommand{\CHBX}{CH~B$^{2}\Sigma^{-}$--X$^{2}\Pi$}

\newcommand{\CNBX}{CN~B$^{2}\Sigma^{+}$--X$^{2}\Sigma^{+}$}

\newcommand{\CaII}{Ca\,{\sc ii}}

\begin{document}
%
%
%

\title{Narrow-band Imaging in the CN Band at 388.33\,nm}
\author{Han Uitenbroek \and Alexandra Tritschler}
\institute{National Solar Observatory/Sacramento Peak\thanks{Operated by the %
           Association of Universities for Research in Astronomy, Inc. (AURA), %
           for the National Science Foundation}, P.O.\ Box 62, Sunspot,
           NM 88349, U.S.A.
           \\ \email{huitenbroek@nso.edu, atritschler@nso.edu}}
\date{\today}


\abstract
{}
{We promote the use of narrow-band (0.05 --- 0.20\,nm FWHM)
  imaging in the molecular
  CN band head at 388.33\,nm as an effective method for monitoring
  small-scale magnetic field elements because it renders them with
  exceptionally high contrast.}
{We create synthetic narrow-band CN filtergrams from
  spectra computed from a three-dimensional snapshot of a magnetohydrodynamic
  simulation of the solar convection to illustrate the expected high contrast
  and explain its nature.
  In addition, we performed observations with the horizontal slit spectrograph at the
  Dunn Solar Tower at 388.3\,nm to experimentally confirm the high
  bright-point contrast, and to characterize and optimize the transmission
  profile of a narrow-band (0.04 FWHM) Lyot filter,
  which was built by Lyot and tailored to the CN band at Sacramento Peak
  in the early 70's.}
{The presented theoretical computations predict that bright-point contrast in
  narrow-band (0.04 FWHM) CN filtergrams is more than 3 times higher than
  in CN filtergrams taken with 1\,nm FWHM wide filters, and in typical G-band
  filtergrams.
  Images taken through the Lyot filter after optimizing its passband confirm
  that the filter is capable of rendering small-scale magnetic elements
  with contrasts that are much higher than in traditional G-band imaging.
  The filter will be available as an user instrument at the Dunn Solar Tower.}
{}


\maketitle           

%
%

\section{Introduction}\label{sec:introduction}
Many Fraunhofer lines weaken in locations where strong small-scale
magnetic fields appear outside sunspots and pores.
While this weakening manifests itself as gaps in the spectral lines,
it produces a local brightening in spectroheliograms produced from
line core intensities, and hence provides a convenient method for
tracking the small-scale magnetic field without the need to collect
polarimetric signals.
This property of spectral lines is exploited particularly effectively
in G-band imaging, in which typically 1 nm full width at half maximum
(FWHM) filters centered on a region of the spectrum around 430.5 nm
with many lines of the CH molecule are employed.
When the CH lines weaken in magnetic elements, the filter
integrated signal increases markedly, providing a high
contrast with respect to the surrounding surface in the filter image
     \citep{Carlsson+Stein+Nordlund+Scharmer2004,Voegler_etal2005,%
Uitenbroek+Tritschler2006}.

A similar opportunity is provided by the spectral region shortward
of 388.33 nm which contains many lines of the $v = 0 - 0$ \CNBX\
system.
Since the CN band is further to the blue than the G band, it potentially
offers higher spatial resolution because of a smaller Airy disk,
and higher contrast rendering of temperature differences, because of the
wavelength dependence of the Planck function.
Indeed, observations presented by
    \citet{Zakharov+Gandorfer+Solanki+Loefdahl2005}
suggest that filtergrams taken through a 0.8 nm FWHM wide filter
centered at 387.9 nm provide higher contrast (by a factor of 1.4)
in the magnetic elements than through a G-band filter of comparable width
(Note that 
    \citet{Zakharov+Gandorfer+Solanki+Loefdahl2005}
erroneously cite 388.7 nm as the central wavelength of their CN filter.)
Theoretical calculations by
    \citet{Uitenbroek+Tritschler2006},
however, find the contrast to be nearly equal, with the
bright points having a slightly lower (by a factor of 0.96) contrast
on average in synthetic CN filtergrams calculated from a
three-dimensional snapshot of a magnetohydrodynamic simulation of
solar convection.
This issue has yet to be resolved by comparing additional simultaneous
observations in both molecular band passes, and modeling in additional
simulation snapshots.

In the present paper we argue that a more appropriate way of
utilizing the weakening of CN lines is by imaging the solar surface
in the band head proper, rather than through a 1 nm type filter.
The band head at 388.33 provides this opportunity because many
lines accumulate there and meld into a broad spectral
feature, thus allowing a for passband that is much broader than a
single CN line.
In addition, the signal of a filtergram through a narrow passband
centered on 388.3 nm, just blueward of the CN band head, is not
affected much by either Doppler shifts in and out of the passband,
or by Zeeman splitting of the CN lines.

The usefulness of imaging the solar surface in the CN band head
was recognized early on, e.g. \citet{gillespie1971}, \citet{sheeley1971},
but the passband became impractical when CCD detectors with much less
sensitivity in the blue started to replace film.
Since fast back-illuminated CCDs with sufficient blue sensitivity are
becoming more prevalent it is time to investigate
the sensitivity of the CN band head intensity to magnetic elements
in more detail.
In this paper we discuss our efforts to revive an old narrow-band Lyot 
filter for the CN band head, characterize its transmission
profile, and present images taken through the filter at the Dunn Solar
Telescope (DST) of the National Solar Observatory in Sacramento Peak. 
The filter has survived mothballing for more than 15 years, appears
in good state and will be available as an user instrument at the DST.
With this effort we hope to make once again available narrow-band
imaging in the CN band as a routine technique for studying the
behavior of small-scale magnetic field elements at the highest spatial
resolution, and with the highest contrast.

This paper is organized as follows.
In Section~\ref{sec:model} we briefly describe the employed method
for generating theoretical synthetic narrow-band filtergrams in the CN
band, and make plausible that these have very high contrast in small-scale
magnetic features.
In Section~\ref{sec:filter} we describe the  narrow-band Lyot filter
for 388.3~nm.
Observations and the characterization of the filter are presented in
Section~\ref{sec:observations}, and conclusions in
Section~\ref{sec:conclusions}.

%
%

\section{Modeling narrow-band CN filtergrams}\label{sec:model}
We explore the benefits of narrow-band CN imaging
in the 388.33 band head with numerical radiative transfer
modeling in a snapshot from a three-dimensional magnetohydrodynamic
simulation of solar surface convection.

\subsection{Method}
\begin{figure*}
  \centering
  \includegraphics[width=17cm]{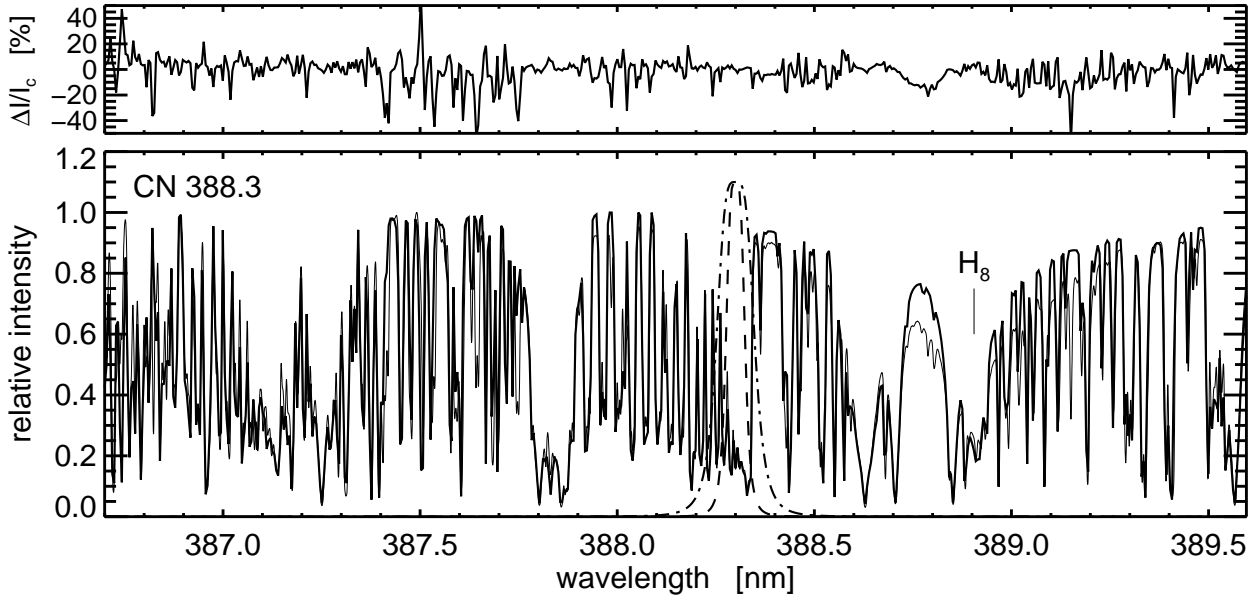}
  \caption{Comparison of the synthetic spectrum over whole computed range,
           averaged over the whole
           surface of the simulation cube, and the observed solar average
           spectrum. Filter curves for 0.05 and 0.10 nm FWHM passbands
           centered at 388.3 nm are also shown.}
  \label{fig:CNspectrum}
\end{figure*}
The method we employ is described in
     \citet{Uitenbroek+Tritschler2006}.
The emergent spectrum was calculated at
600 wavelength positions over the range of 386.7 nm to 389.6 nm
through a three-dimensional simulation snapshot from 
a magnetohydrodynamic simulation of solar convection
     \citep{Stein+Nordlund1998}.
This spectral region
includes 327 lines of the \CNBX\ system ($v = 0 - 0$, where $v$ is
the vibrational quantum number), 231 weak lines ($gf \leq -5$) of the
\CHAX\ system ($v = 0 - 1$ and $v = 1 - 2$),
62 stronger lines of the \CHBX\ system ($v = 0 - 0$), and
the hydrogen Balmer line H$_8$ between levels
$n = 8$ and 2 at $\lambda = 388.905$\,nm.
We use the same snapshot as before.
The emergent spectrum was integrated over filter curves
of different widths and position to create synthetic filtergrams.
The very good agreement between the mean spectrum, averaged over the
surface of the simulation cube, and the observed solar average spectrum
(Figure~\ref{fig:CNspectrum}) indicates that the simulation and
radiative transfer modeling are highly realistic.
The biggest discrepancy between synthetic and observed spectra occurs
in the wings of the Balmer H$_8$ line the wings of which are too narrow
in the former because no linear Stark broadening was included in the
computation.
\begin{figure}[t]
  \centering
  \includegraphics[width=0.66\columnwidth]{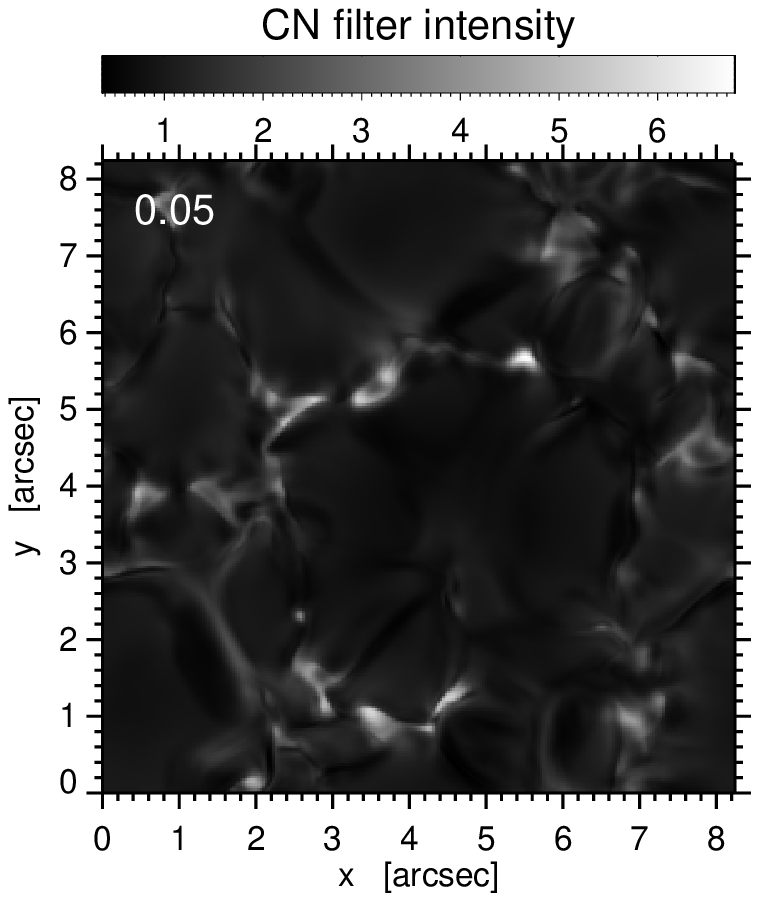}\hfill
  \includegraphics[width=0.66\columnwidth]{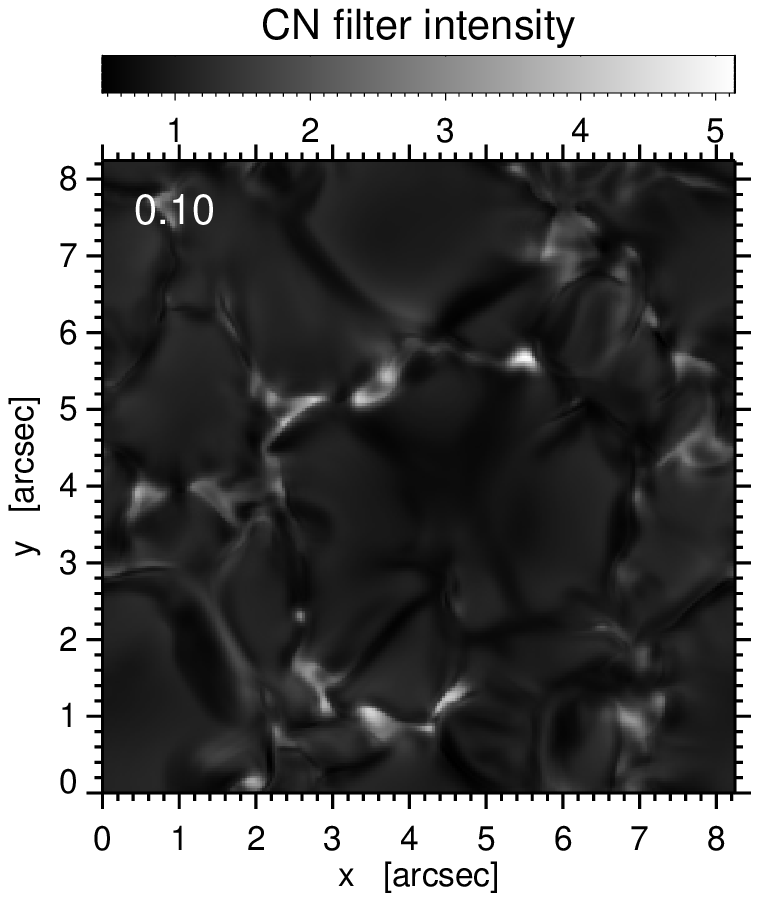}\hfill
  \includegraphics[width=0.66\columnwidth]{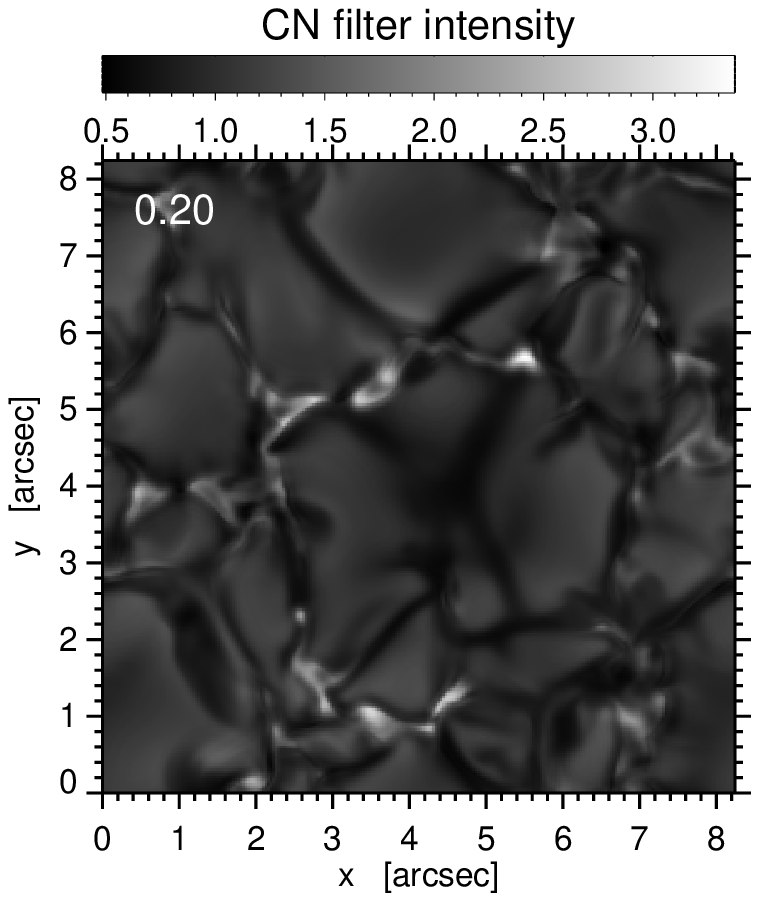}
  \caption{Synthetic CN-band filtergrams for filters
           with FWHM of 0.05\,nm (\textsl{top panel})
           0.10\,nm (\textsl{middle panel}), and 0.20\,nm
           (\textsl{bottom panel}),
           all three centred on the CN band head at 388.3\,nm.}
  \label{fig:CNgrams}
\end{figure}
Magnetic concentrations in the resulting filtergrams, shown in
Figure~\ref{fig:CNgrams}, stand out as bright ribbon-like features.
The average bright-point contrasts for filtergrams synthesized
through 1 nm FWHM G-band and CN-band filter curves, and through
three narrow-band filters of 0.05, 0.10, and 0.20 nm FWHM are
given in Table~\ref{tab:contrasts}.
On average the pixels selected as bright points in the 0.05 nm
and 0.10 nm filters, respectively, have more than three and two
times higher contrast than in the 1 nm FWHM filtergram,
while the 0.20 nm filter only gives an enhancement that is 26\%
higher.
The average bright-point contrast in the synthetic filtergrams
was derived in the same manner as described in
     \citet{Uitenbroek+Tritschler2006},
namely by subtracting a certain fraction of a blue continuum
filtergram from the synthetic G-band filtergram to eliminate the
granulation pattern as much as possible, and then keeping all pixels
above a certain threshold in the difference filtergram.
\begin{table}
  \caption{\label{tab:contrasts} Average bright-point contrasts
           for the G band and wide- and narrow-band CN filter curves}
  \begin{tabular}{lccc}\hline
    Filter  &  CWL [nm]  &  FWHM [nm] &  BP contrast \\\hline
  G band    &  430.5     &   1.00     &   0.497\\
  CN 388.3  &  388.3     &   1.00     &   0.478\\
  CN 388.3  &  388.3     &   0.05     &   1.741\\
  CN 388.3  &  388.3     &   0.10     &   1.150\\
  CN 388.3  &  388.3     &   0.20     &   0.602\\\hline
  \end{tabular}
\end{table}

\subsection{Explanation for the high narrow-band contrast\label{sec:understanding}}
Using a response function of the filter-integrated intensity
to temperature
     \citep{Uitenbroek+Tritschler2006}
pointed out that the brightness enhancement of the magnetic features
in 1 nm FWHM filtergrams in the G band and 388.3 nm CN band does not
result from relatively higher temperatures compared to the surrounding
photosphere, but from the partial
evacuation of the regions of highest magnetic flux concentration.
To balance the sum of magnetic and gas pressures inside a magnetic
field concentration with the gas pressure of the non-magnetic
surroundings the internal gas pressure is substantially lower than
the external one at equal geometric height. In the lower pressure
environment the density of molecules is reduced resulting
in a weakening of the molecular spectral lines for lines of sight that
look down into strong magnetic field concentrations. The weakening
of spectral lines in the filter passband causes the integrated
filter intensity to be higher, leading to the characteristic intensity
enhancements of magnetic elements.
This brightening-through-evacuation mechanism is wavelength
independent resulting in the similar contrasts in the two molecular
bands.

In the CN band head, just longward of 388.3 nm, however, many CN
lines accumulate. The overlap of these multiple lines causes
the formation of intensity over the whole wavelength interval
of the band head to appear in higher layers, where it samples
higher temperatures, at equal geometric height, than in the surrounding
non-magnetic medium because of the shallow temperature gradient
in the flux concentration
     \citep[see][their Figure 9]{Uitenbroek+Tritschler2006}.
This shallow gradient is the result of inhibition of convective
heat flux below the photosphere, and lateral radiative influx
just above the photosphere.
In similar fashion as most photospheric spectral lines weaken,
the raised temperature at the formation height of the lines
raises the central intensity of the CN lines in the band head,
with additional line weakening due to increased molecular dissociation. 
Since the line weakening is a direct measure for the temperature
at the formation height it is advantageous to observe it at
shorter wavelengths, where the response of the Planck function
to temperature is more non-linear.
The advantage of the CN band head is that it is a relatively broad
feature that requires less narrow filters than a single spectral
line, before its filter-integrated intensity becomes contaminated
by continuum contributions.
The latter only starts to occur for filters of 0.20 nm FWHM, or wider,
as is clear in Figure~\ref{fig:CNgrams} and Table~\ref{tab:contrasts}.

%
%

\section{Description of the narrow-band Lyot filter}\label{sec:filter}
We do not know much about the origin of the filter, there seems to
be no documentation
or reports about it. The filter survived languishing in the vaults
of the Evans facility at Sacramento Peak. 
The black metal cover makes it look very similar to an old H$\alpha$
filter that resides also in one of the cabinets at the Evans facility. The 
face plate of the H$\alpha$ filter indicates by a serial number 
and a label that it was built by Bernard Lyot himself. Although this face 
plate is missing for the CN filter we take the similar looks
of both filters as an indication that the CN filter also was built by
Lyot himself but modified at Sacramento Peak probably by Richard Dunn
in the early 70's.
Because of missing documentation we have no insight into
the \emph{inner life} of the filter in terms how many elements (calcites,
polarizers) are used, 
what the length of the thinnest calcite is, etc., which would be necessary
to know in order
to model the transmission band. Removing the outer metal cover
reveals that the optics is wrapped in cork for protection and to insulate
the optics. The thermal control consists of a controller and heating wires
underneath the cork wrap, and an additional temperature sensor with an external
readout device. 
Changing the temperature results in a change of central wavelength (CWL)
and the shape of the passband: 
when heated the filter passband moves towards the blue. 
The entrance polarizer is removable. This disables the thickest element,
effectively doubling the width of the passband
     \citep[e.g, see][]{Stix2004}.
The entrance polarizer can be rotated to align the optical
axis of the entrance polarizer and the first calcite element: fast and slow axis of
the calcite must be 45$^{\circ}$ with respect to the entrance polarizer.
Rotation of the entrance polarizer (if present) also rotates the exit polarizer.

%
%

\section{Observations}\label{sec:observations}
Between 2006 May 1 and May 6, we performed spectroscopic and
imaging observations with the Lyot filter at the Dunn Solar Telescope
(DST) of the National Solar Observatory at Sacramento Peak in order
to characterize its transmission profiles, and to test its
image quality, respectively.
We describe these observations below.

\subsection{Spectroscopic}
The spectroscopic observations presented here have been performed 
between May 1 and May 4 and were done in two different manners.
First, we obtained data to characterize the Lyot filter and its prefilter.
Second, we scanned the solar surface by stepping the Horizontal Spectrograph
(HSG) in order to generate two-dimensional maps at each wavelength position
of the available CN spectrum around 388.33\,nm. In the former case the Lyot
was placed in front of the slit of the HSG.
In both cases the prefilter, needed to eliminate side bands of the Lyot,
was mounted directly in front of the camera.
The available prefilter was an old interference filter, the passband of which
had unfortunately drifted towards the blue (see Sect.~\ref{sec:prefilter}).

Data acquisition was accomplished with a
1024$\times$1024\,pixel$^2$ detector manufactured by Spectral Instruments 
(SI 805-205) with a 13\,$\mu$m pixel size. The chip is backside 
illuminated for enhanced blue sensitivity and has a quantum efficiency 
of $\sim$70\,\% at the observed wavelength. The integration time was set to 3\,s. 
The dispersion in wavelength direction is 10.1\,m\AA\,pixel$^{-1}$ and the
spatial scale along the slit is 0.171\,arcsec\,pixel$^{-1}$.  
\begin{figure}
  \centering
  \resizebox{\hsize}{!}{\includegraphics[]{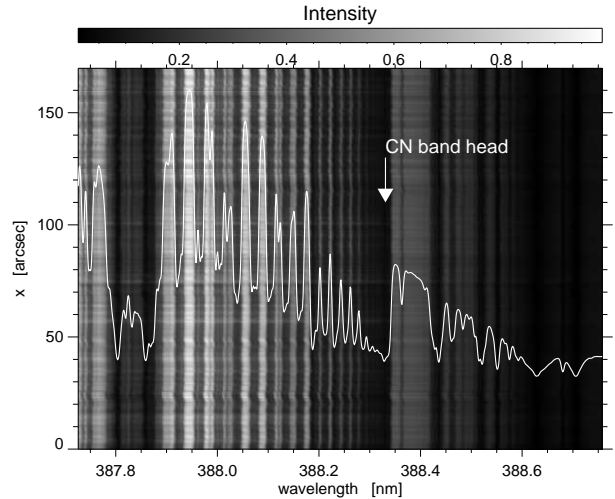}}
  \caption{Sample spectrum of the CN band at 388.33\,nm taken with the prefilter only.
           Overplotted is the averaged spectrum. The CN band head is indicated.
           The spectrum is not corrected for the prefilter transmission curve.}
           \label{fig:spectrum}
\end{figure}
Figure \ref{fig:spectrum} shows a spectrum of the wavelength region
as seen through the prefilter only. The location of the band head is
indicated by the white arrow. 

\subsection{Characterization of the Lyot passband}
\begin{figure}
  \resizebox{\hsize}{!}{\includegraphics[]{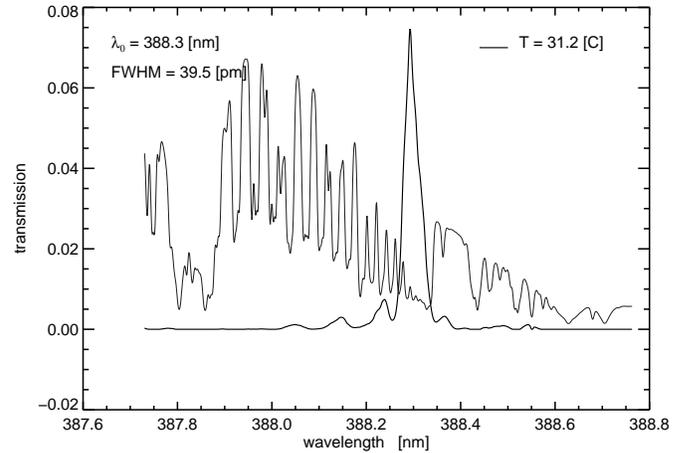}}
  \caption{Optimal passband shape of the CN Lyot filter (\emph{thick curve}).
           With a CWL of 388.3\,nm, just to the blue of the band head at 388.33\,nm
           the width is 0.04\,nm (FWHM). Also plotted is the observed
           spectrum through the prefilter averaged over 32
           frames in flatfield mode.\label{fig:optimal}}
\end{figure}

To characterize the Lyot filter we recorded data with and without the
Lyot in the optical path.
In both cases the telescope was run in flatfield mode, moving the pointing
randomly over the central part of the solar disk during exposures in order
to blur spatial details.
Taking the ratio of the average of the 32 spectra spectra each with and
without the Lyot allows us to determine the passband of the filter. 
We obtained the passband in this manner for different temperature
settings, ranging from 30.4 $^{\circ}$C to 33.2 $^{\circ}$C.
We found that the passband moves towards
the blue with increasing temperature, and splits into two peaks.
Interpolating the measured central wavelength versus temperature
relation we determined that the temperature setting in which the passband
optimally covers the band head without the entrance polarizer present is 31.2 $^{\circ}$C.
The passband for this optimal temperature is shown in Figure~\ref{fig:optimal},
together with the average spectrum obtained by averaging the 32 disk center
spectra taken without the Lyot, but including the prefilter.
Fitting a Gaussian curve to the filter transmission shows that the central
wavelength is 388.3 nm and the width of the filter is 0.04 nm FWHM.
Note that the peak transmission of the filter is almost 8\%, confirming
its excellent optical condition.

\subsection{Passband of the Lyot with entrance polarizer}
\begin{figure}
  \resizebox{\hsize}{!}{\includegraphics[]{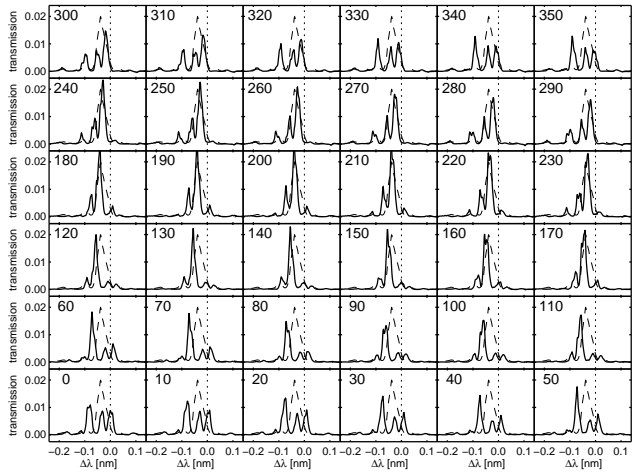}}
  \caption{Variation of the Lyot's passband with rotation angle of 
           the exit polarizing element when the entrance polarizing element
           is mounted (\emph{thick solid curves}),
           which means that the passband is $\sim$0.02\,nm.
           For comparison we overplot the optimal position of the 0.04\,nm passband 
           (entrance polarizing element not mounted, \emph{dashed curve}).
           Wavelength differences are defined with respect to the band head.}
  \label{fig:rotation}
\end{figure}

We also tested the behavior of the Lyot's passband in its narrow setting with
the entrance polarizer in place.
In particular, we obtained the transmission profile of the filter
for different angles of the polarizer, at the optimal temperature
setting for the wide passband configuration.
Rotating the exit polarizer also rotates the entrance polarizer
(and vice versa). 
The result is shown in Figure \ref{fig:rotation} for rotation angles from 0$^{\circ}$ 
to 350$^{\circ}$ in steps of 10$^{\circ}$.
The optimal position of the 0.04\,nm passband is displayed for comparison.  
The best match between the two passbands at a temperature of 31.2 C would be at an
angle of 200$^{\circ}$ or 210$^{\circ}$, except for that the narrower passband shows
significant side lobes for most of the angles.
The cleanest narrow passband occurs for 140--150$^{\circ}$, but is shifted
too far to the blue.
Perhaps the latter could be corrected by lowering the operating temperature
slightly, but the much reduced peak transmission of only 2\% makes this configuration
less attractive because of the long exposure times it requires.
The wider passband is narrow enough to profit from the contrast properties of the
of the CN band head feature, and therefore much more attractive in practice.

\subsection{Characterization of the prefilter\label{sec:prefilter}}
\begin{figure}[t]
  \resizebox{\hsize}{!}{\includegraphics[width=0.75\textwidth]{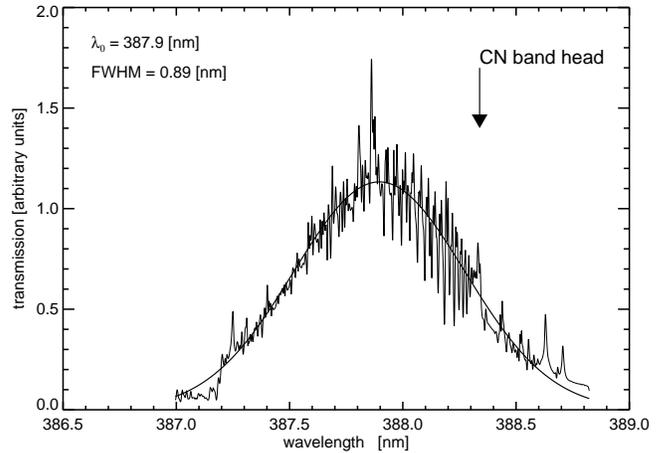}}
  \caption{Ratio of the average observed spectrum with the FTS
           atlas spectrum (\emph{thin curve} representative of the
           prefilter transmission profile. the thick curve is a Gaussian fit
           to the ratio.}
  \label{fig:prefilter}
\end{figure}

For prefilter characterization we removed the Lyot from the 
optical path and took spectral data with the prefilter placed directly 
in front of the detector.
We took the spatially averaged observed spectrum and divided it by the
observed disk center intensity from the FTS atlas by
     \citet{Brault1978,Brault1993} as provided by 
Kurucz\footnote{\textsf{http://kurucz.cfa.harvard.edu/sun/KPNOPRELIM}}.
This ratio, which is rather uneven because of small wavelength offsets
and differences in spectral broadening between the two,
is plotted with the thin curve in Figure \ref{fig:prefilter}.
To get a proper estimate of the CWL and width
of the prefilter's band pass we fitted a Gaussian curve to the
ratio (\emph{thick curve} in  Figure \ref{fig:prefilter}), and found values
of 387.9\,nm and 0.89\,nm (FWHM), respectively.
From this it is clear that the CWL is, unfortunately, too far in the
blue and not centered on the CN band head, and
therefore, makes exposure times longer than necessary.
Tilting the prefilter would not alleviate this problem
as it moves the passband even further towards the blue.

\subsection{Two-dimensional spectral maps\label{sec:heliograms}}
One possible way to test the image quality of the filter is to
construct CN band head spectroheliograms by integrating
observed spectra over the determined filter passband,
and comparing these to imaging observations through the filter.
To this end we generated two-dimensional maps
by scanning the spectrograph slit over the solar surface.
This was done without the Lyot but with the prefilter in the beam.
We scanned areas of 50\,arcsec in steps of 0.25\,arcsec for a total of 200 steps. 
The spectra were taken with an exposure time of 3\,s and a cadence of 5\,s.
The same wavelength range and dispersion was used as for the observations
characterizing the Lyot and prefilter.

\begin{figure}[t]
  \resizebox{\hsize}{!}{\includegraphics[width=0.75\textwidth]{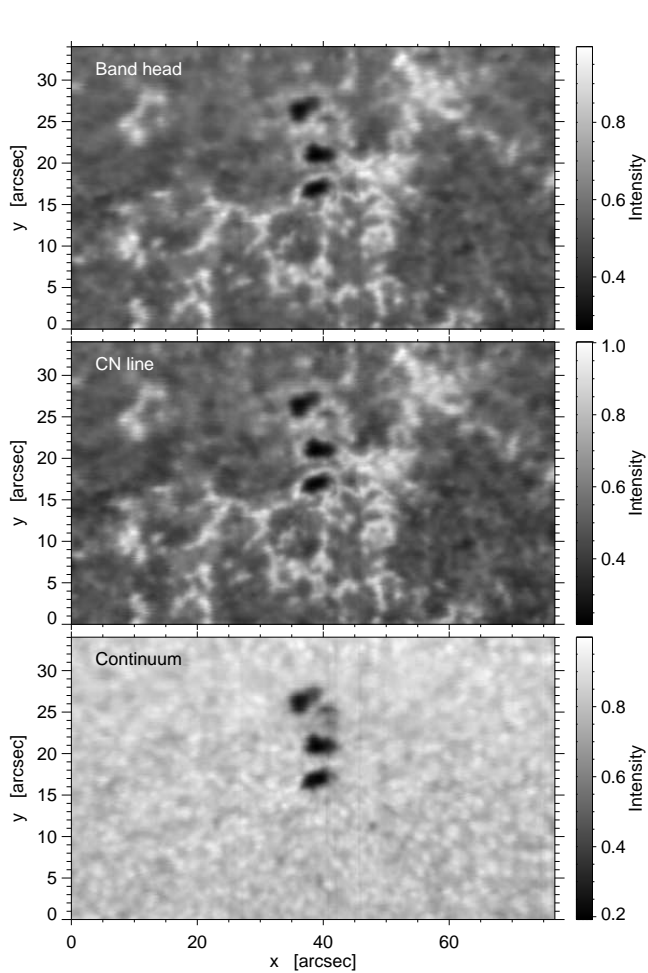}}
  \caption{Area maps of active region NOAA 10879 at N\,15.2 and E\,15.9 observed
           on May 4, 2006. Top: in the CN band head. Middle: in the line core of one
           of the CN lines at 388.10\,nm. Bottom: in the continuum at 387.945\,nm.
           The intensity is scaled to the maximum intensity in the map at
           that wavelength.}
  \label{fig:maps}
\end{figure}
On May 2nd, 2006, we scanned across active region NOAA 10879 at N\,14.9 and E\,10.6. 
(the region was assigned a number only on the next day). 
On May 4th, 2006, we scanned across the same active region now at N\,15.2 and E\,15.9. 
Figure \ref{fig:maps} shows area maps from May 4th for three
different wavelength integrations: the CN band head integrated over the
passband of the Lyot filter (\emph{top panel}), 
the core of the CN line at 388.10\,nm (\emph{middle}), 
and the continuum at 387.945\,nm (\emph{bottom}).
The top two panels clearly show the equivalence of the intensity information
obtained in the core of a single CN line and in the integrated band head.
It is only the width of the latter that makes it much more practical
for filtergram observations.
Since we do not have co-temporal G-band images for the spectral scans
we cannot construct a mask from such filtergrams to compute the
average bright-point contrast.
However, if we assume that the theoretical prediction that the
average bright-point contrast is almost the same in 1 nm CN
band filtergrams as in the G-band, then we can integrate the
observed spectra over a 1 nm FWHM filter function, and derive
a mask from the resulting spectroheliogram.
We find that the average CN bright-point contrast in a 1 nm FWHM
filter calculated in this manner is 0.14, while that through
the optimal passband of the Lyot is 0.38, almost three times
as high as through the wider filter, and in reasonable agreement
with the predicted ratio (see Table~\ref{tab:contrasts}).
The observed contrasts are much lower than the predicted ones
because of image blurring by unfavorable seeing, beyond what could
be corrected by the Adaptive Optics system of the DST.


\subsection{Narrow-band filtergrams}
We observed region NOAA 10879 at N\,15.6 W\,1.6.
on May 4th, 2006 with the filter in an imaging setup.
The exposure time was set to 1\,s with the
same detector as for the spectroscopic observations.
The prefilter was placed in front of the Lyot filter, 
which itself was placed in an F/36 beam with an intermediate
focal plane just after the exit window of the Lyot.
Two reimaging lenses were used to bring the final image
scale to 0.0474\,arcsec\,pixel$^{-1}$, which is\
equivalent to 3.65\,arcsec\,mm$^{-1}$ an almost identical
to that in the telescope's prime focus.
This resulted in a field-of-view (FOV) of 48$\times$48\,arcsec.
\begin{figure}[t]
  \resizebox{\hsize}{!}{\includegraphics[width=0.75\textwidth]{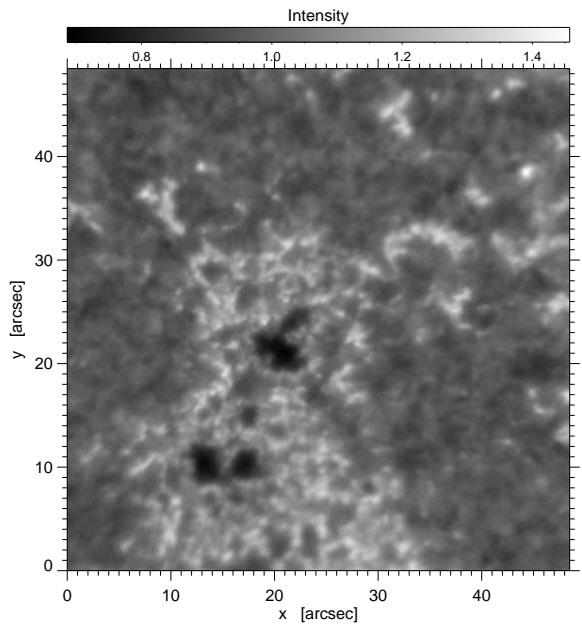}}
  \caption{Image taken centered on the CN rotation band at 388.33\,nm with
           a passband width of $\sim$0.04\,nm (FWHM).
           The temperature was set to 31.2 $^{\circ}$C.}
  \label{fig:image}
\end{figure}

Note the close similarity between the filtergram in 
Figure~\ref{fig:image} and the heliograms in the CN band head and
CN line intensity (Figure~\ref{fig:maps}, \emph{top} and \emph{middle panels},
respectively).
We also note the resemblence with filtergrams in the \CaII\ H and
K lines
     \citep[e.g., ][their Figure 2]{Rouppe_etal2005}, which represent,
however a slightly higher layer of the atmosphere.
In both cases the contrast of small-scale magnetic elements is much
higher than in the 1\,nm wide G-band filtergrams that are
typically used to identify flux concentrations, 
and allows us to use CN filtergrams as a better proxy for
observing small-scale magnetic flux concentrations even when the seeing
conditions are not in favor.

%
%

\section{Conclusions}\label{sec:conclusions}
Given the very similar contrast in the two molecular bands at
430.5\,nm and 388.3\,nm, it would seem logical to conclude that
imaging in the CH band is still the preferred method to track
small-scale magnetic fields because of higher detector sensitivity at
the longer wavelength.
Experimenting with the width and central wavelength of the filter
function in the simulations, however, we confirm theoretically
that the contrast 
in the CN band can be greatly enhanced by making the filter narrow
(0.05 -- 0.2\,nm), and centering it directly in the band head at
388.33\,nm, a fact known experimentally already in the early
seventies.
Examples of synthetic filtergrams made with narrow filters of widths
0.05\,nm to 0.20\,nm shown in Figure~\ref{fig:CNgrams} and the
corresponding average bright-point contrasts listed in Table
\ref{tab:contrasts} illustrate this finding.

The simulations also allow us to understand the high contrasts
in the narrow-band filtergrams, as outlined in
Section~\ref{sec:understanding}.
In the band head the emergent intensity originates from slightly higher
(by about 100-200\,km) layers and ``feels'' the temperature
enhancement that results from lateral irradiation into the slender
flux tubes.
This observing mode therefore favors short wavelengths, because it
preferentially brings out the higher temperature contrast.
The properties of the intensity signal in the CN band head are not
different that those in the core of a single CN line (as illustrated
in Figure~\ref{fig:maps}), observing in the band head is easier
to accomplish because it does not require an extremely narrow filter.

\begin{figure}[t]
  \resizebox{\hsize}{!}{\includegraphics[width=0.65\textwidth]{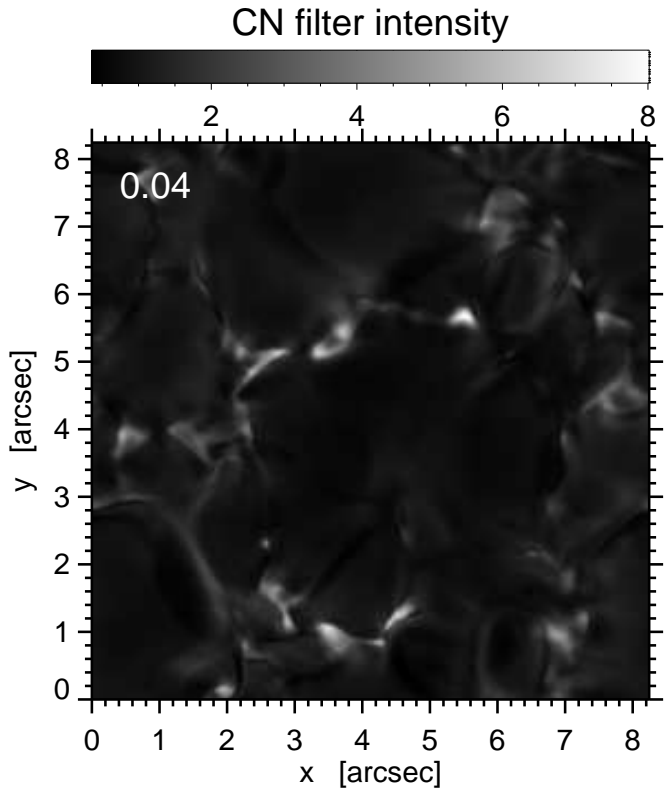}}
  \caption{Filtergram constructed from the theoretical spectrum using
           the measured filter profile (Figure~\ref{fig:optimal})
           at the optimal temperature of 31.2 $^{\circ}$C.}
  \label{fig:optimalmap}
\end{figure}
We tested an old Lyot filter suitable for observations in the
CN band head and determined its passband with spectroscopic
observations at the DST of the NSO at Sacramento peak in May 2006.
We found the filter in very good optical condition,
and determined that its passband at optimal setting has a
width of 39.5 pm (FWHM) and a peak transmission of 7.5\%
(Figure~\ref{fig:optimal}).
To achieve this setting the filter has to be operated at an
internal temperature of 31.2\,C and without its entrance
polarizer, disabling the first birefringent element.
Theoretically, by constructing synthetic filtergrams from 
a magneto-hydrodynamic simulation snapshots with the optimal
filter shape, we expect very high contrasts of small-scale
magnetic features.
Figure~\ref{fig:optimalmap} shows such a filtergram.
The average bright-point contrast computed with a bright-point mask
derived from G-band brightness as described in Section~\ref{sec:model}
is 1.937.

A sample image taken through the Lyot filter is shown in
Figure~\ref{fig:image}.
Because this image was taken under very moderate seeing conditions,
and a fairly long exposure time, necessary because of the off-band prefilter
(see Section~\ref{sec:prefilter}), the contrast is not very high.
Yet, because of the very intrinsic contrast of the bright points they
are distinctly visible in the narrow-band CN image, while they would be
very hard to detect with a wider passband of 1 nm, or through a typical
G-band filter.
Comparison of spectro-heliograms constructed with different filter widths
from our observed spectra confirm our theoretical findings that a 0.05\,nm
FWHM filter centered at 388.3\,nm provides approximately a threefold
enhancement of bright-point contrast over a 1 nm FWHM filter in either
the CN band or the G band (Section~\ref{sec:heliograms}).

We conclude that the Lyot filter is a very valuable instrument that is 
in surprisingly good condition. 
To increase the throughput and shorten exposure times a new prefilter
centered on 388.3\,nm has been ordered.
With this in place the filter will be available as an user instrument
at the DST, and will reenable exploitation of the CN band as a valuable
window for monitoring small-scale magnetic field structure and evolution.

%
%

\acknowledgements
The authors would like to thank the observers, technicians and engineers who made
these observations
possible. Particularly Doug Gilliam who remembered the existence of the filter and thus
prevented it from further languishing in the vaults of the Evans facility. 

%
%


\begin{thebibliography}{11}
\expandafter\ifx\csname natexlab\endcsname\relax\def\natexlab#1{#1}\fi

\bibitem[{{Brault}(1978)}]{Brault1978}
{Brault}, J.~W. 1978, in Future solar optical observations needs and
  constraints, 33

\bibitem[{{Brault}(1993)}]{Brault1993}
{Brault}, J.~W. 1993, Solar Fourier transform spectroscopy (in Future Solar
  Optical Obervations Needs, and Constraints 1978) (Selected Papers on
  Instrumentation in Astronomy), 273

\bibitem[{{Carlsson} {et~al.}(2004){Carlsson}, {Stein}, {Nordlund}, \&
  {Scharmer}}]{Carlsson+Stein+Nordlund+Scharmer2004}
{Carlsson}, M., {Stein}, R.~F., {Nordlund}, {\AA}., \& {Scharmer}, G.~B. 2004,
  \apj, 610, L137

\bibitem[{{Gillespie}(1971)}]{gillespie1971}
{Gillespie}, B. 1971, \solphys, 21, 93

\bibitem[{{Rouppe van der Voort} {et~al.}(2005){Rouppe van der Voort},
  {Hansteen}, {Carlsson}, {Fossum}, {Marthinussen}, {van Noort}, \&
  {Berger}}]{Rouppe_etal2005}
{Rouppe van der Voort}, L.~H.~M., {Hansteen}, V.~H., {Carlsson}, M., {et~al.}
  2005, \aap, 435, 327

\bibitem[{{Sheeley}(1971)}]{sheeley1971}
{Sheeley}, N.~R. 1971, \solphys, 20, 19

\bibitem[{{Stein} \& {Nordlund}(1998)}]{Stein+Nordlund1998}
{Stein}, R.~F. \& {Nordlund}, {\AA}. 1998, \apj, 499, 914

\bibitem[{{Stix}(2004)}]{Stix2004}
{Stix}, M. 2004, The Sun: an introduction (Astronomy and astrophysics library,
  Berlin: Springer, 2004. ISBN: 3540207414)

\bibitem[{{Uitenbroek} \& {Tritschler}(2006)}]{Uitenbroek+Tritschler2006}
{Uitenbroek}, H. \& {Tritschler}, A. 2006, \apj, 639, 525

\bibitem[{{V{\" o}gler} {et~al.}(2005){V{\" o}gler}, {Shelyag}, {Sch{\"
  u}ssler}, {Cattaneo}, {Emonet}, \& {Linde}}]{Voegler_etal2005}
{V{\" o}gler}, A., {Shelyag}, S., {Sch{\" u}ssler}, M., {et~al.} 2005, \aap,
  429, 335

\bibitem[{{Zakharov} {et~al.}(2005){Zakharov}, {Gandorfer}, {Solanki}, \&
  {L{\"o}fdahl}}]{Zakharov+Gandorfer+Solanki+Loefdahl2005}
{Zakharov}, V., {Gandorfer}, A., {Solanki}, S.~K., \& {L{\"o}fdahl}, M. 2005,
  \aap, 437, L43
\end{thebibliography}

\end{document}